\documentstyle[prl,aps,epsf,multicol,psfig,graphics,amsfonts]{revtex}
\begin{document}
\draft
\title{Global phase diagram of bilayer quantum Hall ferromagnets}
\author{M. Abolfath$^{1,2}$, L. Radzihovsky$^3$, and A.H. MacDonald$^2$}
\address{$^1$Department of Physics and Astronomy, University of
  Oklahoma, Norman, OK 73019}
\address{$^2$Department of Physics, University of Texas, Austin, TX 78712}
\address{$^3$Department of Physics, University of Colorado, Boulder, CO 80309}
\date{\today}
\maketitle

\begin{abstract}
  
  We present a microscopic study of the interlayer spacing $d$ versus
  in-plane magnetic field $B_\parallel$ phase diagram for bilayer
  quantum Hall (QH) pseudo-ferromagnets.  In addition to the
  interlayer charge balanced commensurate and incommensurate states
  analyzed previously, we address the corresponding 
  interlayer charge unbalanced ``canted'' QH states.  We predict a 
  large anomaly in the bilayer capacitance at the canting transition
  and the formation of dipole stripe domains with periods exceeding 
  1 micron in the canted state.

\end{abstract}

\begin{multicols}{2}
\narrowtext 

There now exists considerable
experimental~\cite{eisenstein92,murphy94,spielman00,spielman01,muraki01}
and
theoretical~\cite{fertig89,macdonald90,wen_zee92,yang_moon94,tunneling01}
evidence for interlayer phase coherent states in bilayer quantum Hall
(QH) systems at total Landau level filling fraction $\nu_T=1$.  The
ground state in these systems can be regarded as an easy-plane
ferromagnet\cite{yang_moon94}, as a condensate of electrons in one
layer and holes in the Landau level of the other
layer\cite{yoshioka90}, or as a superfluid of Chern-Simons composite
bosons\cite{wen_zee92}.  The competition between tunneling energy,
which pins the interlayer phase, and Coulomb interaction energy, which
favors interlayer phase rigidity, yields a rich phenomenology.  This
is especially true when an in-plane magnetic field component
$B_{\parallel}$, that favors the development of Aharonov-Bohm (AB)
phases, is present.  The nature of the ground state is dependent on
the separation $d$ between layers, on the strength of interlayer
tunneling, which is conventionally parameterized by the splitting
$\Delta_{SAS}$ between symmetric and antisymmetric bilayer
single-particle states, and for $\Delta_{SAS} \ne 0$ on
$B_{\parallel}$.

\begin{figure}[bht]
  \center \epsfxsize 7.7cm
  \vskip-0.7cm
  \rotatebox{-90}{\epsffile{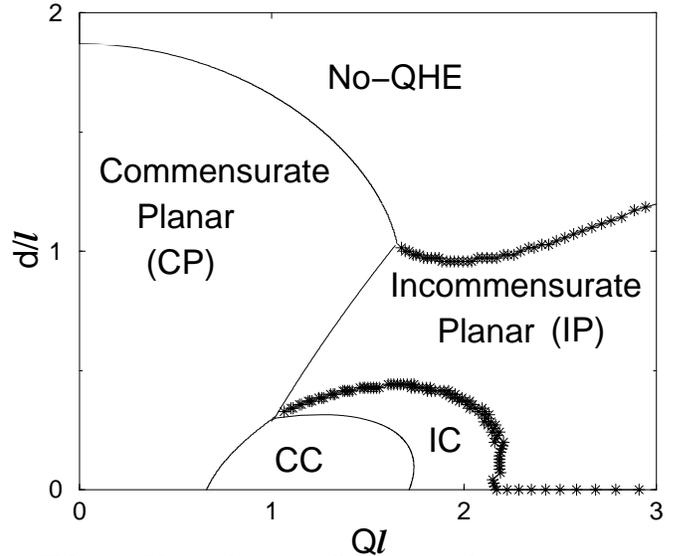}} \vskip-0.3cm 
  \caption{Phase diagram illustrating the commensurate planar (CP),
    incommensurate planar (IP), commensurate canted (CC), and
    incommensurate canted (IC) QH states, for $\Delta_{SAS}=0.126
    e^2/\epsilon \ell$. For smaller $\Delta_{SAS}$, canted and
    incommensurate states shift to smaller $Q$. The nature of the
    compressible ``No-QHE'' state cannot be established on the basis
    of HF theory.}
\label{phase_diagram}
\end{figure}

In this Letter we present a microscopic theory of the
$d-B_{\parallel}$ phase diagram, allowing for the {\em spontaneously}
interlayer charge unbalanced ``canted'' commensurate (C) and
incommensurate (I) QH states proposed recently by one of
us\cite{LRrecent}, in addition to the corresponding charge balanced
``planar'' QH states.\cite{yang_moon94} Our main results are
summarized by the phase diagram of Fig.\ref{phase_diagram}, where five
different phases occur.  In the pseudospin ferromagnet language, the
effect of tunneling is to add to the Hamiltonian an in-plane Zeeman
{\em pseudo}-field, which winds uniformly in space along the $\hat{\bf
  x}$ axis at rate $Q=(B_{\parallel}/B_\perp)(d/\ell^2)$ for a field
in the $\hat y$ direction.  Here the magnetic length $\ell$ is related
to the perpendicular field by $2 \pi \ell^2 B_{\perp} = hc/e$.  In the
{\em planar} charge balanced commensurate (CP) QH state, the
pseudospin magnetization faithfully follows the winding pseudo-field
and its azimuthal angle is given by $\phi({\bf r}) = Qx$.  For
sufficiently large $Q$, however, the cost in exchange energy of this
variation becomes too large and phase-slip solitons are nucleated,
leading to a incommensurate planar (IP) QH ground state, with the
CP-IP transition in the universality class of the well-studied C-I
transition\cite{CItransition}.  Within the Hartree-Fock (HF) theory we
employ, the large $d$ instability, previously studied only at $Q=0$,
is to a Wigner crystal state of the bilayer
system\cite{fertig89,macdonald90}.  It has been argued
previously\cite{macdonald90} that quantum melting of this crystal
leads to a state which does not have broken translational symmetry
but, like the Wigner crystal, is compressible, does not exhibit a QH
effect, and is not distinguished by any symmetry from the state with
uncorrelated $\nu=1/2$ layers that is expected at very large $d$.  We
have followed\cite{note} that suggestion here by labeling this region
``No-QHE''.  Our work focuses on the small $d$ instabilities of the
planar QH pseudospin ferromagnets.
\noindent
 
At small $d$ and intermediate in-plane fields, we find that
$B_\parallel$ drives the easy-plane XY anisotropy\cite{jungwirth} of
these quantum Hall ferromagnets through zero, leading to a continuous
Ising-like {\em reentrant} quantum transitions from the charge {\em
  balanced} CP and IP QH states to the corresponding commensurate (CC)
and incommensurate (IC) interlayer charge {\em unbalanced} canted QH
phases, recently predicted by one of us\cite{LRrecent}.  Our
$d-B_\parallel$ phase diagram (Fig.\ref{phase_diagram}) was
constructed by considering a set of single-Slater-determinant
variational wavefunctions which allow translational symmetry to be
broken in one direction:
\begin{equation}
|\Psi[{\hat{\bf m}}_{X}]\rangle = \prod_{X} (u_X c^{\dagger}_{T,X} + v_X c^{\dagger}_{B,X}) |0\rangle,
\label{varwf}
\end{equation}
where $c^{\dagger}_{T,X}$ and $c^{\dagger}_{B,X}$ create electrons in
top (T) and bottom (B) layers respectively, $X$ is a Landau gauge
guiding center label in the lowest Landau level,
and the pseudospin magnetization ${\hat{\bf m}}_{X} =
(\sin\theta_X\cos\phi_X, \sin\theta_X\sin\phi_X, \cos\theta_X)$ at
label $X$ is the coherent state label of the spinor $(u_X,v_X)$, {\it
i.e.}  $(u_X,v_X) = (\cos[\theta_X/2],
\sin[\theta_X/2]\exp[i\phi_X])$.  
In these wavefunctions $\phi_X$ is the local pseudospin phase coherence
angle and $\theta_X$ is the local polar angle that 
specifies the magnitude of the
charge imbalance between top and bottom layers.  
Taking the expectation value of the microscopic Hamiltonian ${\cal H}$, 
this variational wavefunction leads to the microscopic energy functional
$E_{HF}[{\hat{\bf m}}_{X}]=\langle \Psi[{\hat{\bf m}}_{X}] \vert {\cal
H} \vert \Psi[{\hat{\bf m}}_{X}]\rangle$
\end{multicols}
\widetext
\begin{eqnarray}
E_{HF}[{\hat{\bf m}}_{X}]&=&
-{1\over  2}\Delta_{SAS}\sum_X \sin\theta_X \cos(\phi_X-Q X)
+ {1\over  4 L_y}\sum_{X,X'}\left[2H(X-X')-F_S(X-X')\right] \cos\theta_X \cos\theta_{X'} 
\nonumber\\
&&-{1\over  4 L_y}\sum_{X,X'} F_D(X-X') \sin\theta_X \sin\theta_{X'} 
\cos(\phi_X-\phi_{X'}),
\end{eqnarray}
\begin{multicols}{2}
\columnwidth3.4in
\narrowtext
\noindent
where $\Delta_{SAS}=\Delta_{SAS}^{(0)} \, e^{-Q^2\ell^2/4}$ varies
weakly with in-plane field, and we have assumed a presence of a fixed
neutralizing background of positive charge.  In the above, the terms
proportional to $H(X)$ arise from Hartree (electrostatic)
contributions to the energy functional, while those proportional to
$F_S(X)$ and $F_D(X)$ originate from Fock (exchange) interactions
between electrons in same and different layers respectively.  The
first term in $E_{HF}[{\hat{\bf m}}_{X}]$ is due to interlayer
tunneling and incorporates the AB phases in the factor
$\cos(\phi_X-QX)$.
For a model with arbitrarily narrow two-dimensional electron layers
\begin{mathletters}
\begin{eqnarray}
H(X)&=&\int \frac{dq}{2\pi} \frac{2 \pi e^2 (1 - e^{-qd})}{2 q}
e^{iqX} e^{-q^2\ell^2/2},\label{hofx}\\ 
F_C(X)&=&e^{-X^2/2\ell^2} \int\frac{dq}{2\pi} V_C(q,X/\ell^2) e^{-q^2\ell^2/2},
\label{fofx}
\end{eqnarray}
\end{mathletters}
where $V_C(q_x,q_y) = 2 \pi e^2/q$ and $2\pi e^2 \exp(-qd)/q$ for
$C=S$ and $C=D$ respectively. Note that the exchange integral drops
rapidly with orbit center separation, while the electrostatic integral
falls only as $X^{-2}$ at large $X$, corresponding to interactions
between lines of interlayer charge imbalance electric dipoles. 

The states we discuss are all extrema of this energy functional and
therefore have pseudospin configurations that satisfy the following
two equations:
\end{multicols}
\widetext
\begin{mathletters}
\begin{eqnarray}
\frac{\sin(\phi_X)}{\cos(\phi_X)}&=& 
\frac{\Delta_{SAS} \sin(QX) +{1\over L_y} \sum_{X'} F_D(X-X') \sin(\theta_{X'}) \sin(\phi_{X'})}
{\Delta_{SAS} \cos(QX)+{1\over L_y} \sum_{X'} F_D(X-X') \sin(\theta_{X'}) \cos(\phi_{X'})},\label{phisceq}\\
\frac{\cos(\theta_X)}{\sin(\theta_X)}&=&
\frac{{1\over L_y} \sum_{X'} \cos(\theta_{X'}) [F_S(X-X')-2H(X-X')]}
{\Delta_{SAS}\cos(\phi_X - Q X)
+ {1\over L_y}\sum_{X'} \sin(\theta_{X'}) F_D(X-X') \cos(\phi_{X}-\phi_{X'})}.
\label{thetasceq}
\end{eqnarray}
\end{mathletters}
\begin{multicols}{2}
\columnwidth3.4in
\narrowtext
\noindent
Eqs.\ref{phisceq},\ref{thetasceq} follow from the minimization
of $E_{HF}[{\hat{\bf m}}_{X}]$ with respect to $\phi_X$ and
$\theta_X$, respectively.  The numerator and denominator on the right
hand side of Eq.~\ref{phisceq} are the $\hat x$ and $\hat y$
components of the pseudospin Zeeman effective fields seen by the
Hartree-Fock quasiparticles, which include contributions from
interlayer exchange interactions.  Note that the exchange local
pseudo-field decreases when $\phi_{X}$ is not constant.  Equation
\ref{thetasceq} expresses the property that in the HF ground state the
pseudospin is aligned at each $X$ along the direction of the
pseudospin Zeeman effective field. For example, the commensurate
planar CP state ($\phi_X=QX$ and $\sin(\theta_X) \equiv 1$) solves
these equations with a Zeeman pseudospin field
$\Delta_{SAS}+\tilde{F}_D(Q)$ lying in the xy-plane (with Fourier
transform convention, $\tilde{f}(p) = {1\over L_y} \sum_X f(X)
\exp(-ipX)$).  Because the same layer exchange energy normally
dominates the electrostatic energy, the $\hat z$ component of the
effective pseudo-field tends to have the same sign as the $\hat z$
component of the pseudospin, enabling the canted configurations we
will discuss shortly.

\begin{figure}
\center
\vskip-0.5cm
\epsfxsize7.0cm\rotatebox{-90}{\epsffile{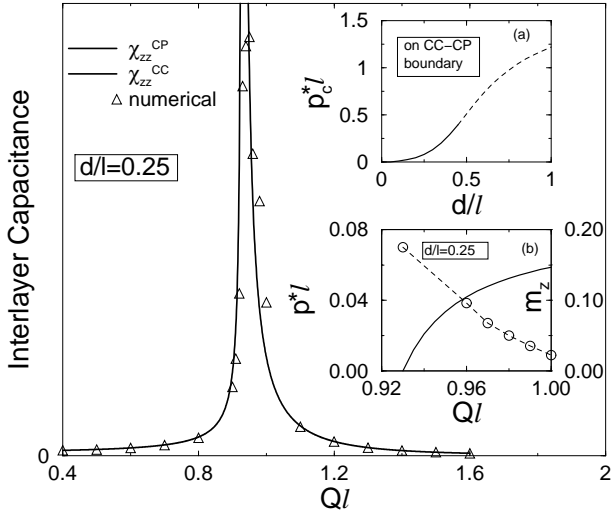}}
\vskip0cm
\caption{
  Bilayer capacitance across CP-CC phase boundary, with peak's height
  set by wavevector $p^*_c$ (inset (a)) of pseudo-magnetization
  $m_z(X)$ that spontaneously develops at the transition.  Inset (b)
  shows the increasing lowest harmonic amplitude of $m_z(X)$ and the
  corresponding wavevector $p^*(Q)$ that decreases inside the CC
  phase.}
\label{capacitance_mz}
\end{figure}

The CP and IP ground states are stable against canting if all
eigenvalues of the following matrix are positive:
\end{multicols}
\widetext
\begin{eqnarray}
K_{zz}(X,X')
&\equiv& 
\frac{1}{2\pi}
\left.{\delta^2 E_{HF}\over \delta m^z_X \delta m^z_{X'}}\right|_{m_z=0} 
= \frac{1}{4\pi L_y} \left[2H(X-X')-F_S(X-X')\right] \nonumber \\ &&
+ \frac{1}{4\pi}\delta_{X,X'} \left[\Delta_{SAS} \cos(\phi_X - QX) 
+ {1\over L_y} \sum_{X'} F_D(X-X') \cos(\phi_X-\phi_{X'})\right].
\label{Kzz}
\end{eqnarray} 
\begin{multicols}{2}
\columnwidth3.4in
\narrowtext
\noindent
In the CP state, translational invariance simplifies the evaluation of
the eigenvalue spectrum of $K_{zz}(X,X')$ which has plane waves
$\exp(ipX)$ eigenfunctions with eigenvalues $\tilde{K}_{zz}(p) =
[\Delta_{SAS} + \tilde{F}_D(Q) + 2 \tilde{H}(p) - \tilde{F}_S(p)]/(4\pi)$.
An important feature of $\tilde{K}_{zz}(p)$ is the non-analytic linear
decrease in $\tilde{H}(p) = (e^2d/2\ell^2)(1 - p d/2 + p^2
(d^2-3\ell^2)/6 + \ldots)$ at small $p$, which originates from the
slow $X^{-2}$ fall off in the (anti-ferroelectric) dipole
electrostatic interactions.  One consequence is that the minimum in
$\tilde{K}_{zz}(p)$ always occurs at a finite $p=p^*_c$, with
$p^*_c\propto d^2$ at small $d$, (see the inset (a) of
Fig.\ref{capacitance_mz}) leading to a minimum at finite $p$.  For a
sufficiently large value of $d$, $K^*\equiv \tilde{K}_{zz}(p^*)$ is
negative even at $Q=0$; this critical value of $d$ is assocated with
the onset of the ``No-QHE'' compressible regime at $Q=0$. For finite
small $Q$, $\tilde{F}_D(Q) = \tilde{F}_D(0)- 4\pi \rho_s (Q\ell)^2 +
\ldots$\ decreases, thereby expanding the ``No QHE'' regime
quadratically with the applied in-plane field $B_\parallel$, as
illustrated in Fig.\ref{phase_diagram}.

The physics of the small $d$ part of the phase diagram is different.
The minimum of $\tilde{K}_{zz}(p)$ ($K^*$) occurs at a finite, but
much smaller value of $p$, and decreases with $Q$, crossing zero
before the incommensurate state boundary is reached as illustrated in
Fig.~\ref{phase_diagram}.  The instability is associated with a change
in sign of the pseudospin anisotropy energy and is closely analogous
to a transition in which the easy axis of a thin-film ferromagnet
changes from in-plane to perpendicular-to-plane, forcing the formation
of stripe domains\cite{domains}.  In the quantum Hall case it is {\em
  electric} rather than {\em magnetic} dipole interactions that force
the transition to occur at finite wavevector.
The CP-CC phase boundary is defined by
$K^*(d,Q)=0$ curve ($d(Q)\approx
(8\pi\rho_s\ell^4/e^2)(Q^2-\xi^{-2})$, rising linearly above the $d=0$
critical value $Q_{CP-CC}=\xi^{-2}$) as illustrated in
Fig.~\ref{phase_diagram}.  As indicated there, the CP-CC canting
instability is preempted at intermediate values of $d$ and $Q$ by the
CP-IP commensurate-incommensurate transition, which for small
$\Delta_{SAS}$ takes place at a small value of $Q=Q_{CP-IP} =
4/(\pi\xi)$, determined by the condition of vanishing soliton energy,
with $\xi = \sqrt{4\pi\ell^2\rho_s/\Delta_{SAS}}$ the width of an
isolated phase slip soliton in the incommensurate
state\cite{CItransition}.  Within the HF approximation  
$Q_{CP-CC}/Q_{CP-IP}$ approaches $\pi/4$ as $d \to 0$.  At finite values of
$\Delta_{SAS}$, the CP-IP phase transition is located by extrema of
the Hartree-Fock energy functional for which $\tilde \phi_X \equiv
\phi_X - QX$ is periodic (modulo $2\pi$), at a $Q$ for which the
period extrapolates to infinity.

Because of its inhomogeneous nature, locating the canting instability
of the IP state is significantly more involved, but conceptually
similar to the analysis of the homogeneous CP state discussed
above.  We first solve the mean-field equations for the IP extremum of
the energy functional, explicitly seeking a self-consistent solution
for $\tilde \phi_X$ with period $a$. This minimization is performed
numerically for a finite value of $L_y$ so that the number of distinct
guiding centers per soliton $N_g = a L_y/(2 \pi \ell^2)$ is finite.

The stability limit of this planar soliton lattice QH state is marked
by the appearance of a zero eigenvalue in a corresponding
$K_{zz}(X,X')$ matrix, which must be computed numerically.
In analogy with the CP-CC transition, here the transition is between
{\em planar} (charge-balanced) IP and {\em canted} (charge-imbalanced)
IC phases, both of which are incommensurate QH states, with IC state
located on the small $d$ side of this phase boundary, as illustrated
in Fig.~\ref{phase_diagram}.  Some features of these non-trivial
numerical results can be understood on the basis of qualitative
considerations.\cite{CItransition} For large $Q$, the AB phases are so
unfavorable for interlayer correlations that the ground state
approaches a simple state in which tunneling between the layers is
ignored completely and $\phi_X$ approaches a constant to minimize the
interaction energy, {\it i.e.}, $\tilde \phi_X = Q X$.  In this limit
the canting stabilities must be the same as those of bilayer systems
with $\Delta_{SAS}=0$. Consequently, as illustrated in
Fig.~\ref{phase_diagram} we find that the canting transition is
{\em reentrant}\cite{LRrecent}, and for large in-plane fields only a
single instability to a ``No-QHE'' state exists, with the critical
value of $d$ universally smaller than that for the $Q=0$ case.  The
evolution of this IP-IC phase boundary with parallel field is
calculated here for the first time, and it would be interesting to
test experimentally.  In the remainder of the paper we concentrate on
the properties of the novel {\em canted} QH states, CC and IC, at
small $d$, which we expect to be reliably rendered by the Hartree-Fock
microscopic theory.

Canted QH phases are distinguished from their planar counterparts by a
finite z-component of the pseudo-spin magnetization order parameter,
$m_z$, which measures the interlayer charge imbalance
$n_{T}-n_{B}=m_z/(2\pi\ell^2)$, that spontaneously develops inside
canted states. As discussed above, because the canting instability is
at a finite wavevector $p^*_c(d)$, for finite $d$ the order parameter
$m^z_X$ is staggered with period $2\pi/p^*_c$. However, in the limit
of small $d\ll\ell$, such that $p^*_c\rightarrow 0$ (or looking at
scales smaller than $2\pi/p^*_c$), $m^z_X$ is nearly uniform and, as
in the magnetic case, it is often a good approximation to ignore the
stripe domain structure and look at the $p=0$ (uniform $\theta_X$)
extrema of the energy functional.  Eq.\ref{thetasceq} then leads to
nonzero $m_z(d,B_\parallel)=\sqrt{1-\sin^2\theta}$, with $\sin\theta =
\Delta_{SAS}/[\tilde{F}_S(0)-2\tilde{H}(0)-\tilde{F}_D(Q)]$, which
therefore predicts the expected mean-field square-root growth of the
canting order parameter (Fig.\ref{capacitance_mz}, inset (b)) inside
the CC phase upon crossing the CP-CC phase boundary in any direction.
Inside the CC state, Eq.\ref{phisceq} predicts the quasi-particle gap
(measured through the activated behavior of the longitudinal
resistivity) to be
$\Delta_{QH}=\Delta_{SAS}+\tilde{F}_D(Q)\sqrt{1-m_z^2(d,B_\parallel)}$,
reduced relative to that of the CP state.
%
\noindent
We are now in the position to calculate the phase boundary for the
CC-IC transition. Since the effect of canting on the soliton physics
is to reduce the one-particle tunneling energy by a factor of
$\sin\theta$ and the interlayer exchange interaction by a factor of
$\sin^2\theta$, standard analysis\cite{CItransition} leads to the
phase boundary
$Q_{CC-IC}(d,B_\parallel)=4/(\pi\xi_0)/\sqrt{\sin\theta}$, that is
shifted to higher critical values of the in-plane field (see
Fig.\ref{phase_diagram}).\cite{LRrecent} We also expect that because
of the long-range (dipole) soliton interaction in the IC state, the
usual\cite{CItransition} $1/|\ln(Q-Q_{CI}|$ rise in the soliton
density will be replaced by a significantly slower
$|Q-Q_{CC-IC}|^{1/2}$ increase inside the IC state.\cite{LRrecent}

In contrast to the CC state, the charge imbalance $m^z_X$ is a
periodic function inside the IC state, even in the $d\rightarrow0$
limit, oscillating with period $a$ of the soliton lattice around the
mean value of the charge imbalance $m^z_0(d,B_\parallel)$. 

The four QH phases that we have discussed are connected by novel
continuous quantum phase transition discussed in Ref.\cite{LRrecent}.
We therefore expect and find a variety of universal experimental
signatures near phase boundaries of Fig.\ref{phase_diagram}. As
illustrated in Fig.\ref{capacitance_mz}, some of many striking
predictions is a strong d-dependent peak in the bilayer capacitance
near the CP-CC phase boundary, as well as the development of
spontaneous interlayer charge imbalance (proportional to $m_z$) inside
the CC phase. We expect, that these and other critical
anomalies\cite{LRrecent}, as well as the dipolar stripe order of CC
and IC phases, with period of order micron and tunable with $d$ and
$B_{||}$, should be readily observable in experiments that we hope our
work will stimulate.

In summary, we have computed $d$ versus $B_\parallel$ phase diagram
for QH bilayers, and found that in addition to the previously studied
planar commensurate and incommensurate phases (which at large $d$ are
unstable to a compressible ``No-QHE''state), there exist interlayer
charge imbalanced commensurate and incommensurate phases, in which
pseudo-magnetization continuously cants out of the easy-xy-plane. We
computed the charge imbalance, differential capacitance and single
particle gap in these new phases, and suggested ways of accessing this
physics experimentally.

This work was supported by the NSF MRSEC DMR-0080054 (MA), Oklahoma
State Regents for Higher Education (MA), NSF DMR-9625111 (LR), the
Sloan and Packard Foundations (LR), the Robert A. Welch Foundation
(AHM), and by the NSF DMR-0115947 (AHM).

\end{multicols}

\begin{references}
\bibitem{comment_dipolar} The situation here is not different from
  ordinary ferro-magnets (electrics), where for sufficiently long
  scales long-range dipolar interactions always dominate the exchange
  and lead to antiferro-magnetic (electric) domain structure.
\end{references}
\end{document}